\documentclass[pdflatex,sn-mathphys-num]{sn-jnl}

\usepackage{graphicx}%
\usepackage{multirow}%
\usepackage{amsmath,amssymb,amsfonts}%
\usepackage{amsthm}%
\usepackage{mathrsfs}%
\usepackage[title]{appendix}%
\usepackage{xcolor}%
\usepackage{textcomp}%
\usepackage{manyfoot}%
\usepackage{booktabs}%
\usepackage{algorithm}%
\usepackage{algorithmicx}%
\usepackage{algpseudocode}%
\usepackage{listings}%

\usepackage{bm}
\usepackage{hyperref}

\newcommand{\diff}{\mathrm{d}}

\theoremstyle{thmstyleone}%

\theoremstyle{thmstyletwo}%

\theoremstyle{thmstylethree}%

\begin{document}

\title[Gravitational-wave imprints of Kerr-BR black holes]{Gravitational-wave imprints of Kerr-Bertotti-Robinson black holes: frequency blue-shift and waveform dephasing}

\author*[1]{\fnm{Xiang-Qian} \sur{Li}}\email{lixiangqian@tyut.edu.cn}
\author[1]{\fnm{Hao-Peng} \sur{Yan}}\email{yanhaopeng@tyut.edu.cn}
\author[1]{\fnm{Xiao-Jun} \sur{Yue}}\email{yuexiaojun@tyut.edu.cn}

\affil*[1]{\orgdiv{College of Physics and Optoelectronic Engineering},
           \orgname{Taiyuan University of Technology},
           \orgaddress{\city{Taiyuan}, \postcode{030024}, \country{China}}}

\abstract{
We investigate the orbital dynamics and gravitational wave signatures of neutral Extreme Mass Ratio Inspirals (EMRIs) in the spacetime of a Kerr black hole immersed in an asymptotically uniform magnetic field, described by the exact Kerr-Bertotti-Robinson (Kerr-BR) solution~\cite{Podolsky:2025tle}. Unlike the widely used Kerr-Melvin metric, the Kerr-BR solution is of algebraic type D, allowing for a rigorous analysis of geodesics and possessing a clear asymptotic structure. By analyzing the Innermost Stable Circular Orbit (ISCO), we confirm that the external magnetic field consistently pushes the ISCO to larger radii. However, contrary to Newtonian intuition, this radial expansion is accompanied by a systematic magnetically induced hardening of the spectrum, where the ISCO frequency is blue-shifted relative to the vacuum case. Notably, in the strong-field regime, we identify a non-monotonic frequency evolution, where the orbital frequency initially decreases before rising rapidly near the horizon, fundamentally altering the chirp character. We further demonstrate that retrograde orbits are significantly more sensitive to magnetic fields than prograde orbits, leading to frequency crossover phenomena where magnetic effects can invert the usual spin-frequency hierarchy. Finally, employing a semi-analytic adiabatic evolution scheme, we quantify the dephasing accumulated during the final year of inspiral. Our results demonstrate that space-borne detectors like LISA can distinguish magnetic environments from vacuum spacetimes for field strengths as low as $B \sim 10^{-4}$, suggesting that environmental magnetic fields could introduce systematic biases in parameter estimation if not properly modeled.
}


\maketitle

\section{Introduction}\label{sec:intro}

The detection of gravitational waves by the LIGO–Virgo–KAGRA collaboration~\cite{LIGOScientific:2018mvr,LIGOScientific:2020ibl,KAGRA:2021vkt} and the imaging of supermassive black holes by the Event Horizon Telescope~\cite{EHT2019,EHT2022} have ushered in an era of precision gravity. While the Kerr metric~\cite{Kerr:1963ud} remains the standard paradigm for describing astrophysical black holes, it strictly represents a vacuum solution. In realistic astrophysical scenarios, however, black holes are inevitably embedded in nontrivial energetic environments. To capture these effects, extensive theoretical efforts have been devoted to constructing exact or approximate solutions of black holes coupled to various matter–energy distributions. Examples include quintessence-like~\cite{Kiselev:2002dx} and Chaplygin-like~\cite{Li:2019lhr,Li:2022csn,Li:2023zfl} phenomenological cosmological fluids, topological defects such as string clouds~\cite{Letelier:1979ej} and global monopoles~\cite{Barriola:1989hx}, as well as scalar fields or boson clouds~\cite{Detweiler:1980uk,Zouros:1979iw} capable of triggering superradiant instabilities. Among these diverse environmental factors, large-scale magnetic fields are of particular astrophysical relevance due to their ubiquity in accretion flows and their crucial role in jet formation mechanisms. Consequently, understanding how such magnetic environments modify the spacetime geometry and the resulting observables is essential for testing general relativity in the strong-field regime.

Historically, the study of magnetized black holes has largely relied on the Kerr–Melvin solution~\cite{Melvin:1963qx,Ernst:1976bsr,Wald:1974np}, generated via the Harrison transformation. While this spacetime has been extensively used, it exhibits interpretational subtleties, particularly regarding its asymptotic behavior: the magnetic field does not decay at infinity, leading to a Melvin-like rather than asymptotically flat exterior and complicating the standard notion of an asymptotic region and associated conserved quantities. More recently, Podolsk\'y and Ovcharenko constructed a new class of exact solutions to the Einstein–Maxwell equations, among which the Kerr–Bertotti–Robinson (Kerr–BR) black hole describes a rotating black hole immersed in an asymptotically uniform magnetic field aligned with the rotation axis~\cite{Podolsky:2025tle}. Crucially, unlike the Kerr–Melvin solution, which is of algebraic type~I, the Kerr–BR spacetime retains Petrov type~D~\cite{Podolsky:2025zlm,Ovcharenko:2025qov}. This algebraic property is of paramount importance, as it implies the existence of hidden symmetries that facilitate the separability of field equations and geodesic motion~\cite{Gray:2025lwy}.

Since its proposal, the Kerr–BR metric has garnered significant attention as a testbed for various astrophysical phenomena. Its optical characteristics have been extensively mapped. Wang \textit{et al.} derived approximate analytical expressions for the photon sphere and investigated the black-hole shadow, quantifying deviations from the Kerr case~\cite{Wang:2025vsx}. These shadow features were further explored by Ali and Ghosh, who demonstrated that the magnetic deformation enlarges the shadow and modifies its oblateness, offering a potential avenue for parameter estimation~\cite{Ali:2025beh}. Complementary studies of the optical properties~\cite{Zeng:2025tji} and strong-field gravitational lensing~\cite{Vachher:2025jsq} have shown that the magnetic parameter leaves distinct imprints on lensing observables, such as image positions and time delays. Beyond geometric optics, the thermodynamic and energetic properties of Kerr–BR black holes have also been scrutinized. Zeng and Wang analyzed energy extraction via magnetic reconnection, finding that while the magnetic field generally impedes extraction efficiency compared to the vacuum Kerr case, it still allows for more efficient extraction than in the Kerr–Melvin background~\cite{Zeng:2025olq}. Theoretical extensions of the metric have also been proposed, including superpositions with Bonnor–Melvin fields~\cite{Astorino:2025lih}, couplings to nonlinear electrodynamics~\cite{Ortaggio:2025sip}, and the inclusion of string clouds~\cite{Ahmed:2025ril}. Potential applications to high-energy astrophysical events, such as short gamma-ray bursts powered by magnetized mergers, have likewise been suggested~\cite{Rueda:2025lgq}.

The orbital dynamics of test particles, which are central to the modeling of extreme mass-ratio inspiral (EMRI) waveforms, have recently seen rigorous development in this background. Zhang and Wei analyzed the kinematics of spinning test particles, revealing that the magnetic field necessitates increased orbital angular momentum to maintain stability~\cite{Zhang:2025ole}. Most relevant to the present work, Wang derived exact, closed-form expressions for the innermost stable circular orbit (ISCO) radii and analytic inspiral trajectories for uncharged particles in the Kerr–BR spacetime~\cite{Wang:2025bjf}. While Wang's work provided the foundational geodesic integration, the specific implications of these modified trajectories for gravitational-wave generation—particularly for the frequency evolution and phase accumulation of EMRIs—remain to be quantified.

In this paper, we bridge the gap between the exact mathematical solutions of the Kerr–BR geodesic equations and their direct observables in gravitational-wave astronomy. Building upon the exact ISCO relations derived in Ref.~\cite{Wang:2025bjf}, we compute the adiabatic evolution of EMRIs in this magnetized background. We show that the presence of the external magnetic field robustly shifts the ISCO frequency to higher values and induces significant dephasing in the emitted waveforms, providing a distinctive observational signature of environmental effects in future space-based gravitational-wave observations.

\section{The Kerr-Bertotti-Robinson metric}\label{sec:metric}

The Kerr-Bertotti-Robinson (Kerr-BR) metric describes a rotating black hole of mass \(m\) and spin parameter \(a\) immersed in an external magnetic field characterized by the parameter \(B\). In coordinates adapted to its algebraic structure, the line element reads~\cite{Podolsky:2025tle}
\begin{align}
\label{eq:metric}
\diff s^2 = \frac{1}{\Omega^2} \bigg[
& -\frac{Q}{\rho^2}\bigl(\diff t - a \sin^2\theta \, \diff\varphi\bigr)^2
+ \frac{\rho^2}{Q}\,\diff r^2
+ \frac{\rho^2}{P}\,\diff\theta^2 \nonumber \\
& + \frac{P}{\rho^2}\sin^2\theta \bigl(a \diff t - (r^2+a^2)\diff\varphi\bigr)^2
\bigg],
\end{align}
where the metric functions are given by
\begin{subequations}
\begin{align}
\rho^2 &= r^2 + a^2 \cos^2\theta, \\
P &= 1 + B^2\left(m^2 \frac{I_2}{I_1^2} - a^2\right)\cos^2\theta, \\
Q &= (1+B^2 r^2)\,\Delta, \\
\Omega^2 &= (1+B^2 r^2) - B^2 \Delta \cos^2\theta,
\end{align}
\end{subequations}
with the modified horizon function \(\Delta\) and auxiliary constants \(I_1, I_2\) defined as
\begin{equation}
\label{eq:delta_r}
\Delta = \left(1 - B^2 m^2 \frac{I_2}{I_1^2}\right) r^2 - 2m\frac{I_2}{I_1} r + a^2,
\end{equation}
\begin{equation}
I_1 = 1 - \frac{1}{2}B^2 a^2, \qquad I_2 = 1 - B^2 a^2.
\end{equation}
Here, the factor \(\Omega^2\) plays the role of a conformal factor. In the limit \(B=0\), we have \(I_1=I_2=1\), \(\Omega^2=1\), and the metric reduces exactly to the standard Kerr solution in Boyer-Lindquist coordinates. In contrast, for \(m=0\) it reduces to the Bertotti-Robinson universe, a direct-product Einstein-Maxwell solution with a uniform electromagnetic field.

\section{Geodesic motion and analytical relations}\label{sec:geodesics}

We consider the orbital dynamics of a test particle with rest mass \(\mu\) confined to the equatorial plane \((\theta = \pi/2)\). The nonvanishing metric components of the Kerr-BR spacetime [Eq.~\eqref{eq:metric}] on this plane take the compact form
\begin{subequations}
\label{eq:metric_components}
\begin{align}
g_{tt} &= -\frac{1}{\Lambda} \left( \frac{\Lambda \Delta}{r^2} - \frac{a^2}{r^2} \right), \\
g_{t\phi} &= \frac{1}{\Lambda} \left( \frac{a \Lambda \Delta}{r^2} - \frac{a(r^2+a^2)}{r^2} \right), \\
g_{\phi\phi} &= \frac{1}{\Lambda} \left( \frac{(r^2+a^2)^2}{r^2} - \frac{a^2 \Lambda \Delta}{r^2} \right),
\end{align}
\end{subequations}
where \(\Lambda(r) \equiv 1 + B^2 r^2\) is the conformal factor restricted to the equatorial plane, and \(\Delta(r)\) is the modified horizon function defined in Eq.~\eqref{eq:delta_r}.

\subsection{Analytical solution for the ISCO radius}

The dynamics of the inspiral are bounded by the innermost stable circular orbit (ISCO). Recently, Wang~\cite{Wang:2025bjf} derived an exact analytical solution for the ISCO radius in the Kerr-BR spacetime. Remarkably, when expressed in terms of the outer and inner horizon radii \(r_\pm\) (the roots of \(\Delta(r)=0\)), the ISCO radius \(r_{\rm ISCO}\) takes a form formally identical to the standard Kerr case:
\begin{equation}
\label{eq:wang_isco}
r_{\rm ISCO} = \frac{r_+ + r_-}{2} \left[ 3 + Z_2 \mp \sqrt{(3-Z_1)(3+Z_1+2Z_2)} \right],
\end{equation}
where the upper (lower) sign corresponds to prograde (retrograde) orbits. The auxiliary functions are defined as
\begin{subequations}
\begin{align}
Z_1 &= 1 + (1-\lambda^2)^{1/3} \Bigl[(1+\lambda)^{1/3} + (1-\lambda)^{1/3}\Bigr], \\
Z_2 &= \sqrt{3\lambda^2 + Z_1^2}, \\
\lambda &= \frac{2\sqrt{r_+ r_-}}{r_+ + r_-}.
\end{align}
\end{subequations}
Equation~\eqref{eq:wang_isco} provides the precise termination point for our inspiral evolution.

\subsection{General circular orbits and frequency}

To describe the adiabatic evolution before reaching the ISCO, we require the orbital frequency \(\Omega_\phi\) and conserved quantities \((\mathcal{E}, \mathcal{L})\) for general circular orbits at any radius \(r > r_{\rm ISCO}\).

The orbital frequency \(\Omega_\phi = \diff\phi/\diff t\) is determined by the extremum condition of the effective potential, \(\partial_r V_{\rm eff} = 0\). For a generic stationary and axisymmetric spacetime, this condition yields a closed-form expression in terms of the radial derivatives of the metric components (\(g'_{\mu\nu} \equiv \partial_r g_{\mu\nu}\))~\cite{Chandrasekhar:1985kt}:
\begin{equation}
\label{eq:omega_exact}
\Omega_\phi(r; B) = \frac{-g'_{t\phi} + \sqrt{(g'_{t\phi})^2 - g'_{tt}\, g'_{\phi\phi}}}{g'_{\phi\phi}}.
\end{equation}
This relation thus gives the exact Keplerian frequency for equatorial circular geodesics in the magnetized background. By substituting Eqs.~\eqref{eq:metric_components} into Eq.~\eqref{eq:omega_exact}, we obtain the precise frequency evolution used in our waveform modeling. In the limit \(B \to 0\), Eq.~\eqref{eq:omega_exact} reduces to the familiar Kerr expression \(\Omega_\phi = \sqrt{m}/(r^{3/2} + a\sqrt{m})\).

The corresponding specific energy \(\mathcal{E}\) and angular momentum \(\mathcal{L}\) for circular orbits follow from the normalization condition \(u^\mu u_\mu = -1\):
\begin{equation}
\mathcal{E} = -\frac{g_{tt} + \Omega_\phi g_{t\phi}}{\mathcal{N}},
\qquad
\mathcal{L} = \frac{g_{t\phi} + \Omega_\phi g_{\phi\phi}}{\mathcal{N}},
\end{equation}
where the normalization factor \(\mathcal{N}\) is defined as
\begin{equation}
\mathcal{N} = \sqrt{-\Bigl(g_{tt} + 2\Omega_\phi g_{t\phi} + \Omega_\phi^2 g_{\phi\phi}\Bigr)}.
\end{equation}

The effective potential \(V_{\rm eff}(r)\) governing the radial motion for a particle with fixed \(\mathcal{E}\) and \(\mathcal{L}\) can then be written as
\begin{equation}
V_{\rm eff}(r; B) = -1 + \frac{\mathcal{E}^2 g_{\phi\phi} + 2\mathcal{E}\mathcal{L} g_{t\phi} + \mathcal{L}^2 g_{tt}}{g_{t\phi}^2 - g_{tt} g_{\phi\phi}}.
\end{equation}

\begin{figure}[t]
    \centering
    \includegraphics[width=0.6\textwidth]{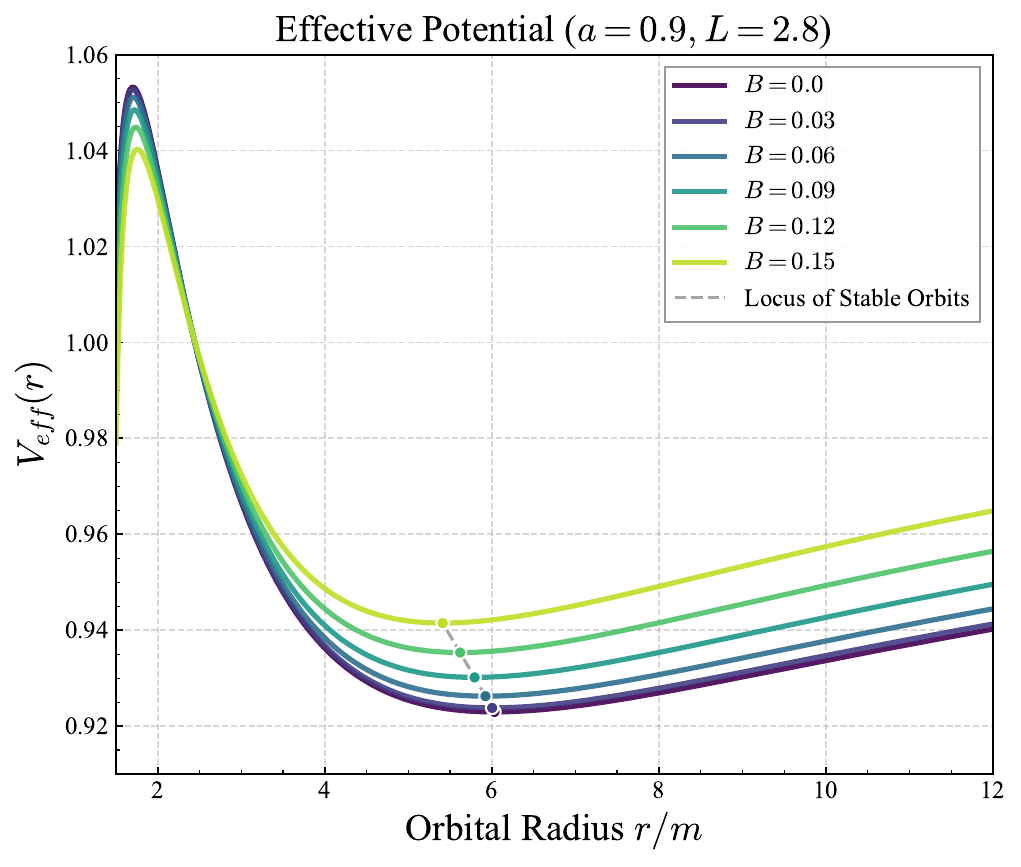}
    \caption{\textbf{Effective potential wells.} The effective potential \(V_{\rm eff}(r)\) for a test particle with fixed angular momentum \(\mathcal{L}=2.8\) around a Kerr-BR black hole with \(a=0.9\). The colored dots mark the locations of stable circular orbits, and the dashed line indicates the locus of these minima as the magnetic field \(B\) increases. Note that for fixed \(\mathcal{L}\), the position of the potential minimum shifts inward as \(B\) increases. The ISCO, however, is defined by the onset of marginal stability \(\partial_r^2 V_{\rm eff}=0\); its behavior as a function of \(B\) is discussed in Sec.~\ref{sec:isco}.}
    \label{fig:potential}
\end{figure}

Figure~\ref{fig:potential} illustrates the behavior of the effective potential for a fixed angular momentum \(\mathcal{L}=2.8\). As the magnetic field \(B\) increases, the potential well deepens and its minimum shifts towards smaller radii, indicating enhanced confinement for particles with fixed \(\mathcal{L}\).

\section{ISCO shift and frequency analysis}\label{sec:isco}
We evaluate the exact analytical ISCO solution derived in Eq.~\eqref{eq:wang_isco} for a range of magnetic field strengths $B$ and spin parameters $a$. Our results reveal a robust yet counterintuitive behavior of magnetized black holes.

\subsection{Outward shift of the ISCO}
\begin{figure}[t]
    \centering
    \includegraphics[width=0.6\textwidth]{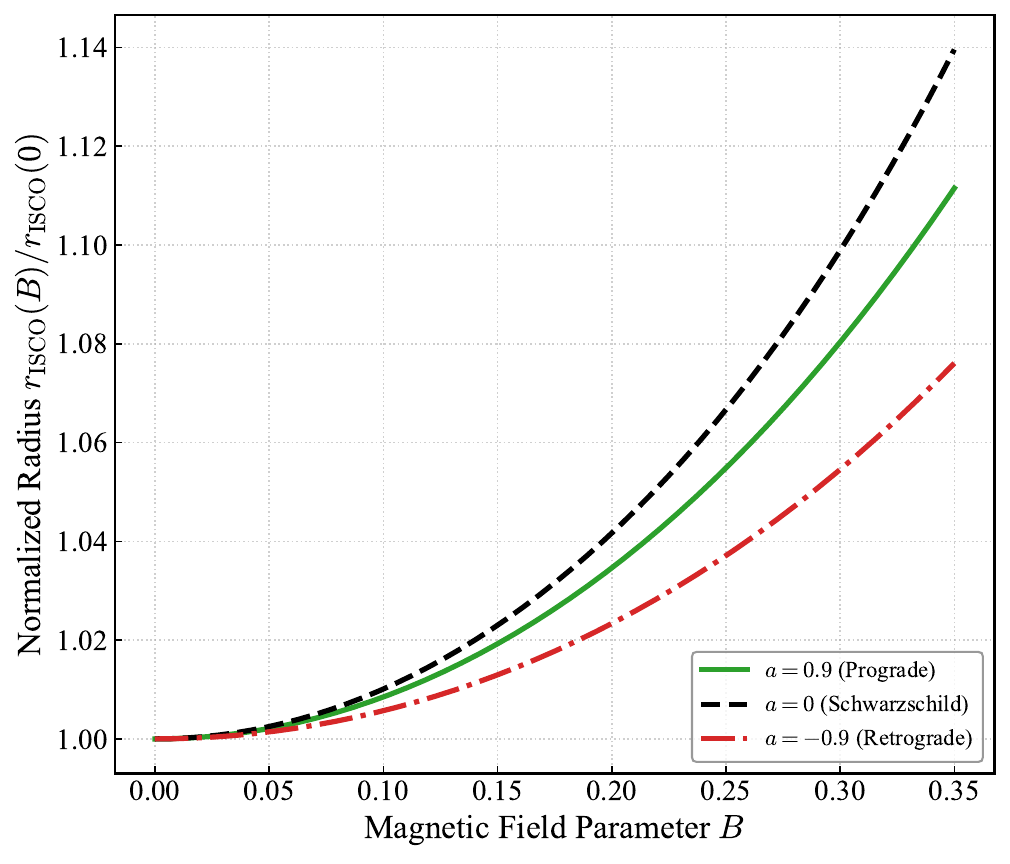}
    \caption{\textbf{Outward shift of the ISCO.} The normalized ISCO radius \(r_{\rm ISCO}(B)/r_{\rm ISCO}(0)\) as a function of the magnetic field parameter \(B\). For all spin orientations, the magnetic field pushes the ISCO to larger radii. The Schwarzschild orbit (black dashed line) exhibits the largest relative sensitivity to the magnetic field, while the high-spin prograde and retrograde orbits show comparatively smaller normalized shifts.}
    \label{fig:isco_radius}
\end{figure}

In Fig.~\ref{fig:isco_radius}, we present the normalized ISCO radius \(r_{\rm ISCO}(B)/r_{\rm ISCO}(0)\) as a function of the magnetic field parameter \(B\). We examine three representative cases: prograde \((a=0.9)\), Schwarzschild \((a=0)\), and retrograde \((a=-0.9)\).

A key finding is that the magnetic field systematically increases the ISCO radius, $r_{\rm ISCO}(B) > r_{\rm ISCO}(0)$, across all spin configurations sampled. This indicates that the uniform magnetic field provides additional radial support that destabilizes the innermost circular orbits and pushes the marginal stability limit outward. Interestingly, the relative shift is most pronounced for the Schwarzschild case ($a=0$), reaching a deviation of over $10\%$ at $B=0.3$. In contrast, both highly prograde and retrograde orbits show a more moderate normalized increase. This suggests that the non-spinning potential well is comparatively ``softer'' against magnetic perturbations relative to its initial radius, whereas the strong frame-dragging effects in high-spin scenarios (whether co- or counter-rotating) impart a rigidity to the spacetime geometry that suppresses the relative radial expansion.

\subsection{Frequency blue-shift and crossover}

Standard intuition from Keplerian dynamics suggests that an orbital expansion leads to a decrease in frequency, $\Omega \sim r^{-3/2}$. In the Kerr-BR spacetime, however, we find the opposite trend at the ISCO.

\begin{figure}[t]
    \centering
    \includegraphics[width=0.6\textwidth]{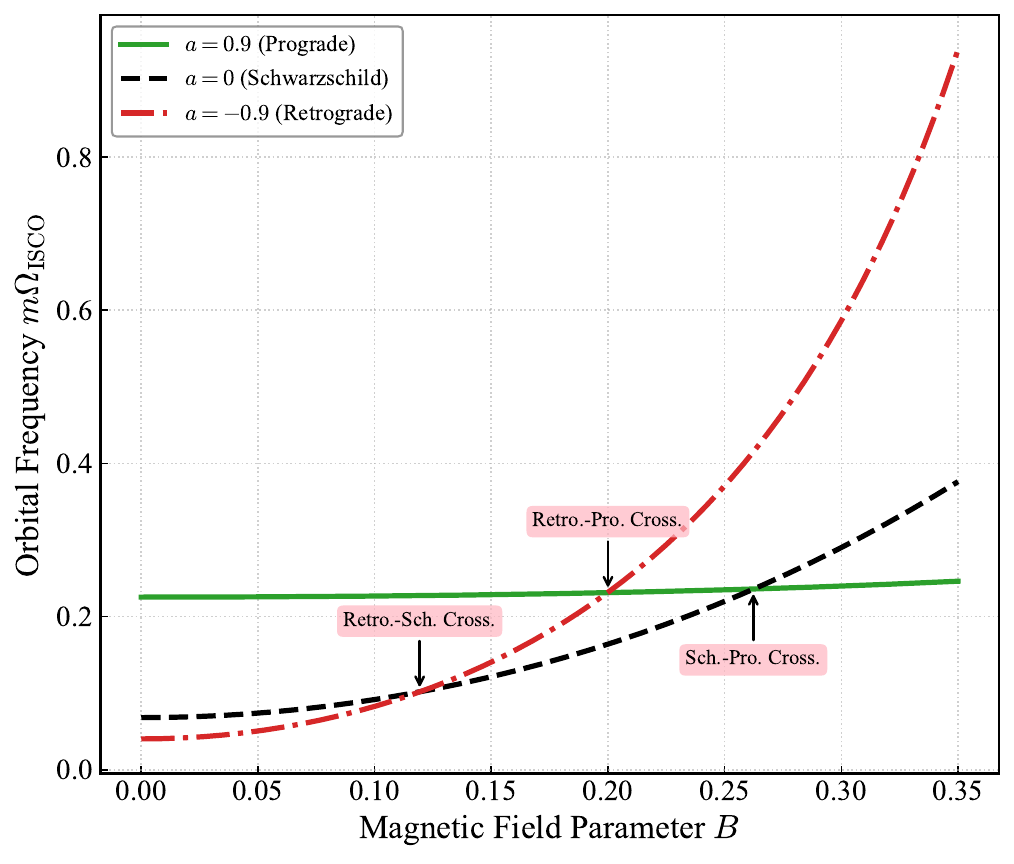}
    \caption{\textbf{Frequency blue-shift and crossovers.} The orbital frequency at the ISCO as a function of $B$. Despite the increase in ISCO radius, the frequency increases for all spin values (blue-shift). Three distinct crossover events are observed: first at $B \approx 0.12$, where the retrograde frequency $(a=-0.9)$ exceeds the Schwarzschild case $(a=0)$; second at $B \approx 0.20$, where the retrograde frequency surpasses the prograde orbit $(a=0.9)$; and third at $B \approx 0.26$, where the Schwarzschild frequency overtakes the prograde orbit. These crossings demonstrate that sufficiently strong magnetic fields can invert the frequency hierarchy typically imposed by the black-hole spin.}
    \label{fig:frequency}
\end{figure}

As shown in Fig.~\ref{fig:frequency}, the ISCO orbital frequency $m\Omega_{\rm ISCO}$ monotonically increases with the magnetic field strength for all spins considered. This ``frequency blue-shift'' implies that the magnetic-field–induced curvature corrections in the Kerr-BR metric force the particle to orbit faster in order to maintain a circular geodesic at the outwardly shifted ISCO. The inset in Fig.~\ref{fig:frequency} confirms that even for high-spin prograde orbits $(a=0.9)$, where the vacuum frequency is already high, the magnetic field induces a further blue-shift.

Most remarkably, we observe a generic ``crossover'' phenomenon. Orbits corresponding to lower spins, which initially have lower ISCO frequencies at $B=0$, exhibit a much steeper rise in frequency as $B$ increases, eventually overtaking the frequencies of higher-spin orbits. Specifically, we identify three crossover events: (1) the retrograde orbit $(a=-0.9)$ surpasses the Schwarzschild case $(a=0)$ at $B \approx 0.12$; (2) at $B \approx 0.20$, the retrograde frequency exceeds even the high-spin prograde orbit $(a=0.9)$; and (3) at $B \approx 0.26$, the Schwarzschild frequency overtakes the prograde one. Thus, beyond a certain field strength, the magnetic corrections completely override the usual spin-imposed ordering of ISCO frequencies.

This counterintuitive behavior and the crossover phenomenon can be understood by examining how the magnetic field couples to the orbital radius. In the Kerr-BR metric, the leading magnetic modification enters through terms scaling as $\sim B^2 r^2$, such as in the conformal factor $\Lambda = 1 + B^2 r^2$. This dependence introduces a fundamental asymmetry in how the magnetic field affects orbits at different radii. The factor $B^2 r^2$ acts as a confining geometrical contribution: for orbits at larger radii this term is significantly larger, implying that the magnetic field produces a stronger effective radial steepening of the potential than it does near the black hole.

To maintain a stable circular orbit against this enhanced radial confinement, the test particle requires a higher angular velocity to provide sufficient centrifugal support. This leads to a differential sensitivity of the frequency to $B$ depending on the ISCO location. Prograde orbits reside deep in the gravitational potential well at small radial coordinates, where the environmental correction factor $B^2 r^2$ remains modest. In contrast, retrograde orbits are located at significantly larger radii (typically nearly an order of magnitude larger than their prograde counterparts). Consequently, in this outer region, the term $B^2 r^2$ is dramatically amplified, subjecting the retrograde orbits to a much stronger magnetic confinement effect. This enhanced sensitivity $(\partial\Omega_{\rm ISCO}/\partial B)$ allows the retrograde frequency to rapidly ``catch up'' and eventually overtake the less sensitive prograde frequency, resulting in the observed frequency crossover.

\subsection{Global parameter-space trends}

\begin{figure}[t]
    \centering
    \includegraphics[width=0.6\textwidth]{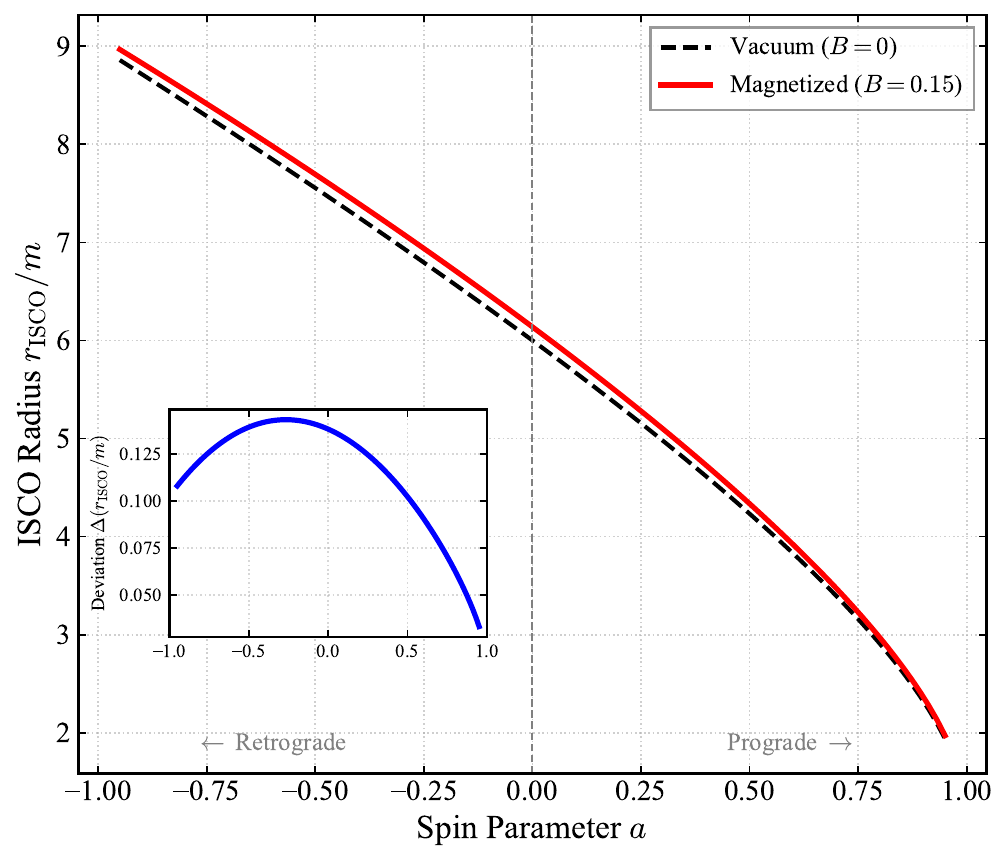}\\[1ex]
    \includegraphics[width=0.6\textwidth]{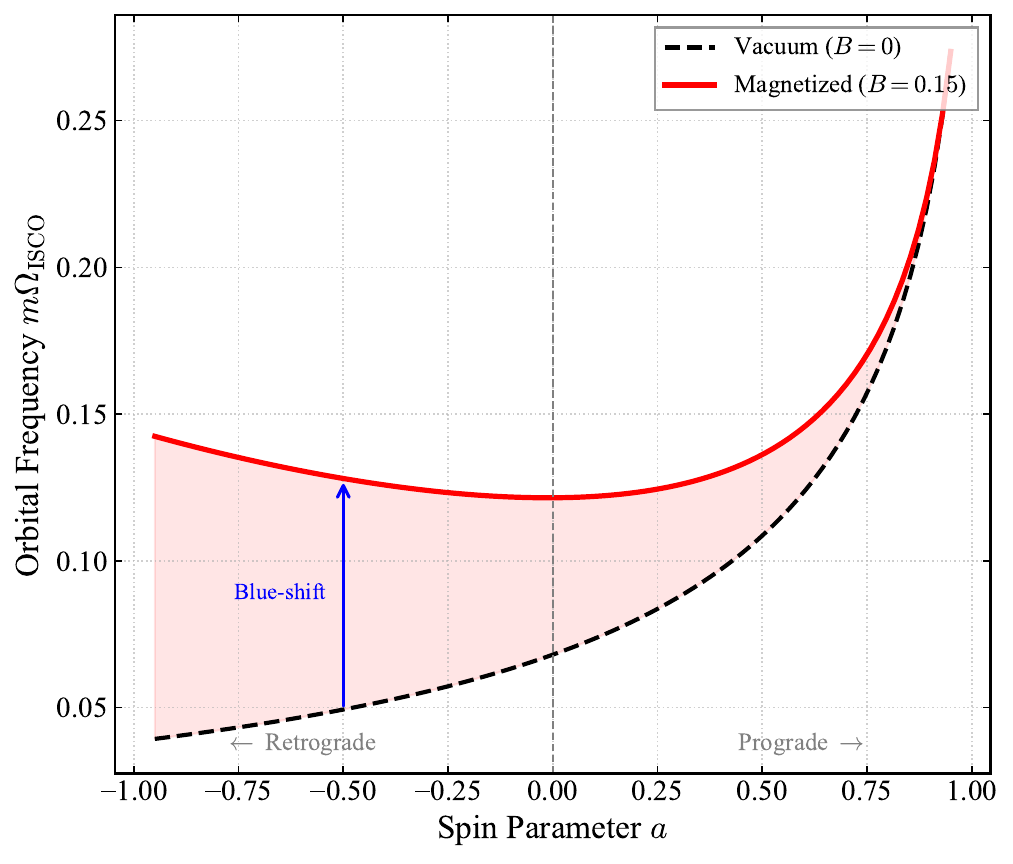}
    \caption{\textbf{Global parameter-space scan.} Top: ISCO radius versus spin $a$ for vacuum (black) and magnetized (red, $B=0.1$) cases. The inset highlights the absolute deviation $\Delta r_{\rm ISCO}$, revealing a nonmonotonic sensitivity that peaks at $a \approx -0.26$. Bottom: ISCO frequency versus spin. The red shaded region represents the magnetic hardening (blue-shift) induced by the field, which is most significant for retrograde orbits.}
    \label{fig:spin_scan}
\end{figure}

To obtain a more global picture of the magnetic imprint, we extend our analysis to the entire spin interval $a \in (-1, 1)$. Figure~\ref{fig:spin_scan} summarizes the interplay between the black hole's intrinsic angular momentum and the external magnetic field.

We first focus on the frequency behavior shown in the bottom panel of Fig.~\ref{fig:spin_scan}. The red shaded region highlights a generic ``magnetic hardening'' of the gravitational-wave spectrum: for any given spin $a$, the magnetized ISCO frequency (red solid line) is higher than the corresponding vacuum value (black dashed line). The separation between the two curves becomes increasingly pronounced in the retrograde regime $(a < 0)$, whereas they tend to merge for rapidly spinning prograde orbits $(a \to 1)$. This behavior confirms that retrograde orbits act as hypersensitive probes of environmental magnetic fields, as the magnetic confinement effect dominates over the comparatively weaker gravitational binding at larger radii.

The top panel of Fig.~\ref{fig:spin_scan} displays the corresponding shift in the ISCO radius. While the magnetic field invariably pushes the ISCO outward $(r_{\rm mag} > r_{\rm vac})$, the magnitude of this shift exhibits a subtle, nonmonotonic dependence on spin, as detailed in the inset. Notably, the absolute deviation $\Delta r_{\rm ISCO} = r_{\rm mag} - r_{\rm vac}$ does not peak at the maximally retrograde spin $(a = -1)$, as one might naively expect from the large orbital radius alone. Instead, the maximum deviation occurs at an intermediate retrograde spin, approximately $a \approx -0.26$.

This nonmonotonic behavior can be qualitatively understood as the outcome of two competing physical mechanisms. First, as $a$ becomes more retrograde, the ISCO radius $r$ increases. Since the magnetic metric corrections scale as $\sim B^2 r^2$, larger radii provide a longer ``magnetic lever arm'' to destabilize the orbit, tending to increase $\Delta r_{\rm ISCO}$. Second, as $a \to -1$, the counter-rotating frame-dragging effects become dominant, creating a ``stiffer'' effective potential well. In this regime, the ISCO location is strongly dictated by the background spacetime geometry, making it increasingly resistant to external magnetic perturbations. The peak at $a \approx -0.26$ therefore represents a balance point between the growing magnetic leverage and the increasing rigidity of the underlying gravitational potential.

\section{Imprint on gravitational waves}\label{sec:gw}

To characterize the imprint of the magnetic field on gravitational waves, we adopt a relativistic adiabatic evolution scheme. In this framework, the conservative sector of the dynamics is governed by the exact geodesic motion in the Kerr-BR metric, so that the nonperturbative effects of the magnetic field on the spacetime geometry are fully retained. The resulting inspiral trajectories provide a direct bridge between the exact mathematical solution and astrophysical observables.

\subsection{Geodesic-driven inspiral evolution}

We model the inspiral phase under the adiabatic approximation, assuming that the radiation--reaction timescale is much longer than the orbital period $(T_{\rm rad} \gg T_{\rm orb})$. The trajectory of the compact object is then determined by the energy-balance equation
\begin{equation}
    \frac{\diff \mathcal{E}_{\rm orbit}}{\diff t} = -\dot{E}_{\rm GW}.
\end{equation}

To maximize the fidelity of the orbital motion, we treat the conservative and dissipative sectors with different levels of approximation. For the conservative sector, we use the exact analytical expressions for the specific energy $\mathcal{E}(r; B)$ and orbital frequency $\Omega_\phi(r; B)$ derived in Sec.~\ref{sec:geodesics}. This ensures that the orbital kinematics fully reflect the magnetically distorted geometry, without invoking weak-field or slow-motion expansions. For the dissipative sector, we employ the leading-order relativistic quadrupole formula for the gravitational-wave energy flux,
\begin{equation}
    \dot{E}_{\rm GW} = \frac{32}{5}\,\eta\,\bigl(m \Omega_\phi(r; B)\bigr)^{10/3},
\end{equation}
where $\eta = \mu/m \ll 1$ is the symmetric mass ratio.

The evolution of the orbital radius is then obtained from
\begin{equation}
    \frac{\diff r}{\diff t}
    = \frac{\diff r}{\diff \mathcal{E}} \frac{\diff \mathcal{E}}{\diff t}
    = - \left( \frac{\partial \mathcal{E}(r; B)}{\partial r} \right)^{-1} \dot{E}_{\rm GW}.
\end{equation}
Here, the radial derivative $\partial \mathcal{E}/\partial r$ is computed symbolically from the metric components [Eqs.~\eqref{eq:metric_components}] rather than by finite differencing, in order to avoid numerical inaccuracies near the ISCO where the potential gradient becomes small.

\subsection{Mechanism of accelerated inspiral and frequency turnover}

\begin{figure}[t]
    \centering
    \includegraphics[width=0.6\textwidth]{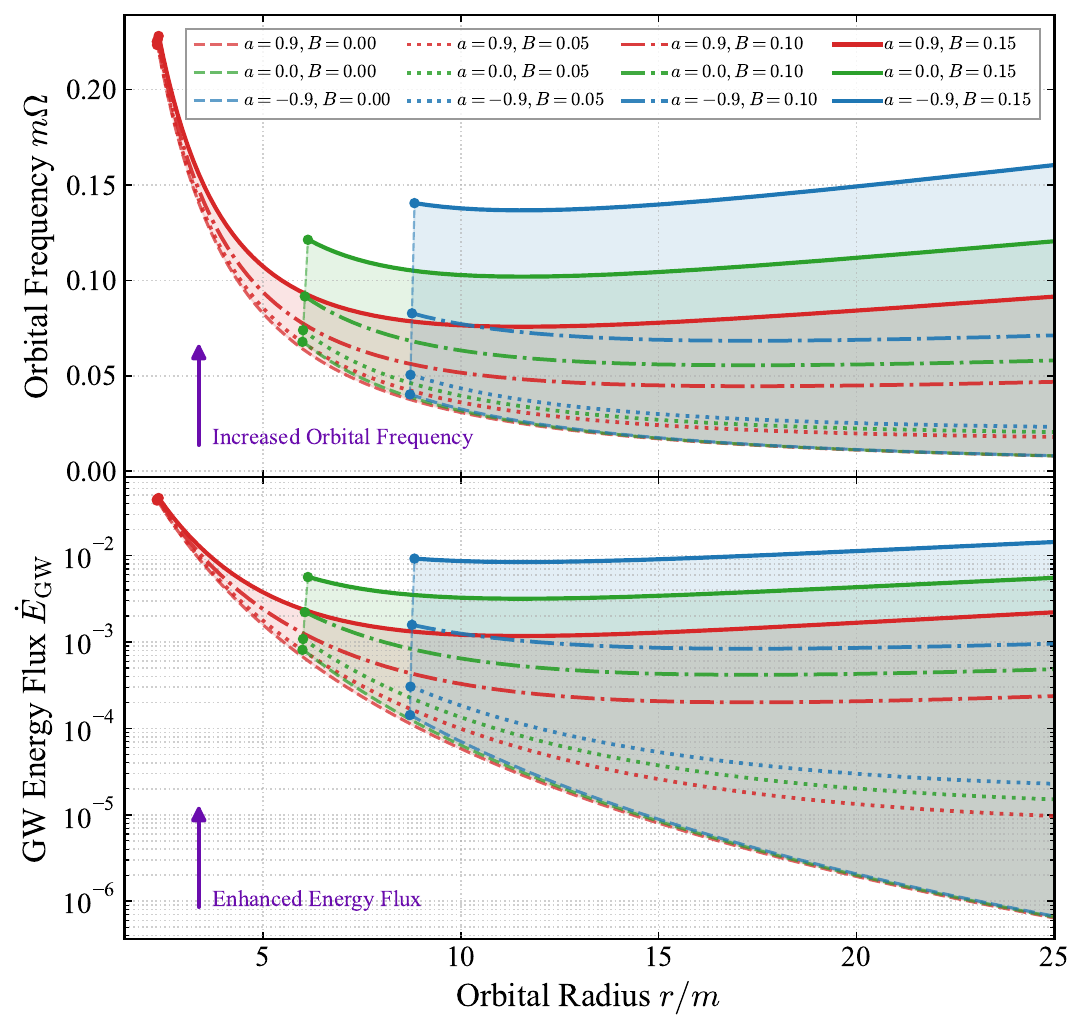}
    \caption{\textbf{Magnetic enhancement of frequency and flux.}
    Evolution of the orbital frequency $m\Omega$ (top panel) and gravitational-wave energy flux $\dot{E}_{\rm GW}$ (bottom panel) as a function of orbital radius $r$ for three spin configurations ($a=0.9, 0, -0.9$) and four magnetic field strengths ($B=0, 0.05, 0.10, 0.15$). The solid circles on the left denote the ISCO location for each configuration.
    For weak fields ($B=0, 0.05$), the frequency increases monotonically as the radius decreases. However, for strong fields ($B=0.10, 0.15$), the frequency exhibits a non-monotonic behavior: it first decreases in the magnetically dominated far region before increasing in the gravity-dominated near region. The shaded regions highlight the range of possible values bounded by the vacuum ($B=0$) and maximal ($B=0.15$) cases, with the left boundary determined by the ISCO trajectory.}
    \label{fig:quantities}
\end{figure}

The physical mechanism driving the orbital evolution is illustrated in Fig.~\ref{fig:quantities}, which displays the radial dependence of the orbital frequency and energy flux for various spin and magnetic field configurations.

First, we observe a robust frequency enhancement in the strong-field regime. As the particle approaches the ISCO, the magnetic field introduces geometrical confinement terms (scaling as $\sim B^2 r^2$) that effectively steepen the radial potential well. To counteract this additional inward curvature and maintain a stable circular orbit, the test particle requires a larger angular velocity than in the vacuum Kerr case. Consequently, near the ISCO, we consistently observe a frequency blue-shift ($\Omega_{\rm mag} > \Omega_{\rm vac}$) for all spins.

Second, a distinct topological transition appears in the global frequency evolution. As shown in the top panel of Fig.~\ref{fig:quantities}, for weak magnetic fields ($B=0$ and $0.05$), the orbital frequency increases monotonically as the radius shrinks from $25m$ to the ISCO. This corresponds to the standard ``chirp'' behavior where the inspiral naturally leads to higher frequencies. However, for stronger fields ($B=0.10$ and $0.15$), the frequency exhibits a non-monotonic behavior: as the orbit shrinks from $25m$, the frequency initially decreases before eventually turning over and increasing rapidly near the ISCO. This implies that in the far region, the magnetic potential dominates and creates an environment where tighter orbits are actually slower---an ``inverse chirp'' regime---before the standard gravity-dominated dynamics take over at small radii.

This dynamical behavior translates directly into the radiative output (bottom panel). Since the flux scales as $\dot{E}_{\rm GW} \propto \Omega_\phi^{10/3}$, the frequency enhancement near the ISCO acts as a nonlinear amplifier, increasing the energy emission rate by orders of magnitude. This generates a feedback loop: the particle radiates energy more efficiently, causing the orbit to shrink more rapidly, which in turn drives the system deeper into the strong-field region.

\subsection{Waveform Dephasing and Cutoff}

We construct a time-domain gravitational-wave strain $h(t)$ using a quadrupole proxy for the dominant $(2,2)$ mode and an optimally oriented observer:
\begin{equation}
    h(t) \propto \bigl(m \Omega_\phi(t)\bigr)^{2/3}
    \cos\!\left[ 2 \int_{0}^{t} \Omega_\phi(t') \,\diff t' \right].
\end{equation}
The evolution is terminated when the particle reaches the ISCO $\bigl(r(t) = r_{\rm ISCO}\bigr)$. We align the waveforms at the end of the inspiral (taken as $t=0$).

\begin{figure}[t]
    \centering
    \includegraphics[width=0.6\textwidth]{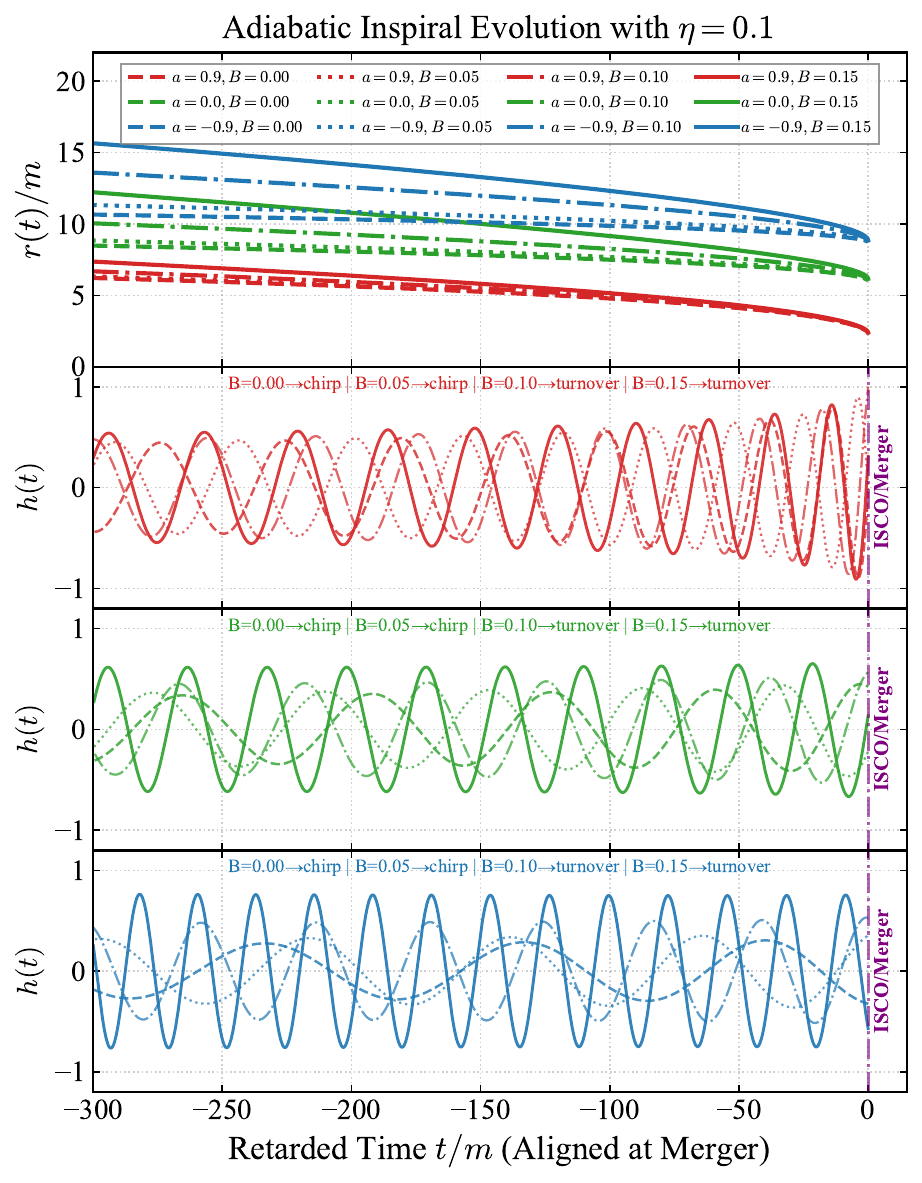}
    \caption{\textbf{Waveform dephasing.}
    Gravitational-wave strains for three spin configurations ($a=0.9, 0, -0.9$) across a range of magnetic field strengths ($B=0, 0.05, 0.10, 0.15$). The text labels indicate the chirp character detected from the frequency evolution $\dot{\Omega}$.
    For all spin configurations, weak fields ($B \le 0.05$) preserve the standard monotonic ``chirp'' behavior ($\dot{\Omega} > 0$).
    However, at higher field strengths ($B \ge 0.10$), the waveforms exhibit a ``turnover'' phenomenon: the frequency first decreases (inverse chirp phase) and then increases (normal chirp phase) as the particle spirals in. This non-monotonic signature is a direct consequence of the competition between the magnetic potential at large radii and the gravitational potential at small radii.}
    \label{fig:waveform}
\end{figure}

The resulting waveforms are shown in Fig.~\ref{fig:waveform}. For visualization purposes, we use an illustrative mass ratio $\eta=0.1$ to display the frequency evolution within a compact timeframe.

The waveform morphology confirms the dynamical features identified in the frequency analysis. For all spin parameters ($a=0.9, 0, -0.9$), when the magnetic field is weak ($B \le 0.05$), the waveforms exhibit the standard ``chirp'' character, where the frequency increases monotonically throughout the inspiral. The primary effect of the magnetic field in this regime is a kinematic acceleration, significantly shortening the inspiral duration compared to the vacuum case.

However, at sufficiently high field strengths ($B=0.10$ and $0.15$), a ``turnover'' phenomenon emerges across all spin configurations. In these cases, the waveform frequency is not monotonic. During the early stage of the inspiral (corresponding to larger radii), the frequency decreases over time, characteristic of an ``inverse chirp'' driven by the dominant magnetic potential $\Lambda = 1+B^2r^2$. As the particle crosses into the strong-gravity regime near the black hole, the frequency evolution reverses, recovering the standard positive chirp $\dot{\Omega} > 0$ before the plunge. This turnover signature---a switch from frequency decrease to increase---uniquely encodes the transition from the asymptotic magnetic environment to the near-horizon gravitational field.

\subsection{Observability of Magnetic Fields}

While the frequency turnover discussed in Sec.~\ref{sec:isco} is a feature of extreme fields, the cumulative phase evolution is sensitive to much weaker, astrophysically realistic magnetic fields. To assess the observability of these effects, we quantify the accumulated dephasing $|\Delta \Phi|$ over a fixed observation window $T_{\rm obs}$ corresponding to the final stage of the inspiral. Specifically, for a typical EMRI system ($\eta=10^{-5}$) detectable by LISA, we define the dephasing accumulated during the last year before the plunge ($t \in [t_{\rm merger}-1\,\text{yr}, t_{\rm merger}]$) as:
\begin{equation}
    \Delta \Phi = \left| \Phi_{\rm mag}(T_{\rm obs}) - \Phi_{\rm vac}(T_{\rm obs}) \right|,
\end{equation}
where $\Phi(t) = \int \Omega(t) dt$ is the phase evolution tracked from a radius $r_{\rm start}$ such that the time to reach the ISCO is exactly one year. This approach ensures that our estimates reflect the actual sensitivity of space-based detectors within a realistic mission lifetime.

\begin{figure}[t]
    \centering
    \includegraphics[width=0.6\textwidth]{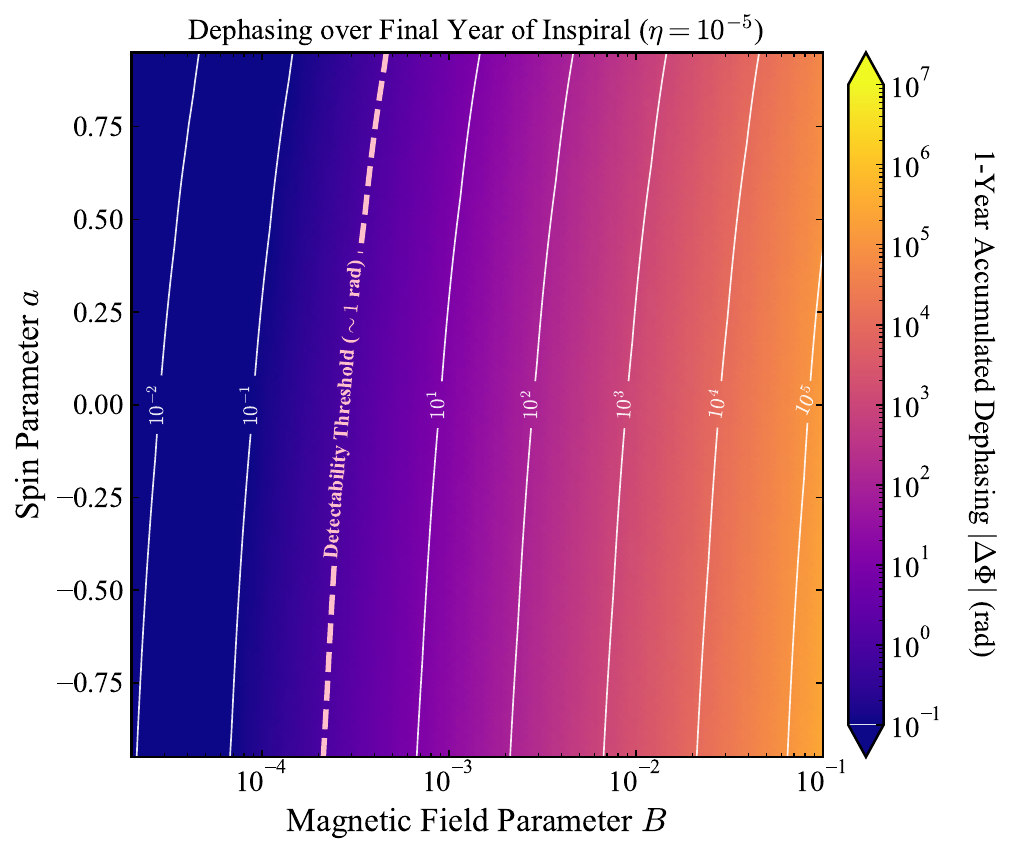}
    \caption{\textbf{Global map of accumulated dephasing over the final year of inspiral.}
    The magnitude of the phase difference $|\Delta \Phi|$ (in radians) accumulated during the last year of observation is shown across the parameter space of magnetic field strength $B$ and spin parameter $a$ for an EMRI with $\eta=10^{-5}$. The color gradient indicates the magnitude on a logarithmic scale. White contours denote orders of magnitude ($10^{-2}$ to $10^{5}$ rad). The pink dashed line marks the conservative detectability threshold ($\sim 1$ rad), which lies in the range $B \sim 2\text{--}4 \times 10^{-4}$. The region to the right of this line represents the parameter space theoretically distinguishable by LISA within a one-year observation window.}
    \label{fig:dephasing_heatmap}
\end{figure}

The global structure of the accumulated phase difference is visualized in the parameter-space heatmap in Fig.~\ref{fig:dephasing_heatmap}, which reveals that the magnetic field strength $B$ acts as the primary driver of the dephasing. The phase difference contours are predominantly vertical, indicating that the magnetic influence largely overshadows the effect of the black hole spin. The color gradient demonstrates a rapid growth in $|\Delta \Phi|$ along the horizontal axis. Even though the instantaneous frequency deviation $|\Omega_{\rm mag} - \Omega_{\rm vac}|$ is minute at the beginning of the observation window (typically $r \sim 20\text{--}30m$), the cumulative phase error integrated over the $\mathcal{O}(10^5)$ orbital cycles typical of EMRIs leads to a significant total dephasing. The pink dashed line marks the conservative detection threshold of $|\Delta \Phi| \sim 1$ rad. In our one-year observation window, this threshold corresponds to a magnetic field strength of approximately $B \sim 2\text{--}4 \times 10^{-4}$. We note that a rigorous distinguishability analysis depends on the signal-to-noise ratio (SNR) $\rho$ of the specific event (typically requiring a mismatch $\mathcal{M} \gtrsim D/2\rho^2$, where $D$ is the number of source parameters). However, assuming a typical EMRI source for LISA with $\rho \sim 20$, a phase shift of unity is widely accepted as a robust necessary condition for detectability~\cite{Zi:2023qfk}. Our results indicate that environmental fields with $B \gtrsim 10^{-4}$ satisfy this condition and are thus potentially observable.

The observability becomes even more pronounced in the strong-field regime. For the astrophysically realistic upper bound of $B \sim 10^{-2}$ suggested in literature~\cite{Wang:2025vsx}, the system enters a high-dephasing regime where the accumulated phase difference reaches magnitudes of $10^3$ rad. In this regime, the magnetic imprint is so dominant that the dephasing crosses the detectability threshold of 1 rad within a timescale of merely hours. This implies that such strong magnetic environments would be unmistakably distinguishable from vacuum spacetimes almost immediately upon detection, regardless of minor variations in detector noise or source orientation.

Beyond the dominant magnetic scaling, the heatmap also reveals a secondary spin dependence. The contours exhibit a systematic curvature to the right as the spin parameter $a$ increases across the entire parameter space. This morphological feature implies that, for a fixed magnetic field strength, retrograde orbits ($a < 0$) accumulate a larger phase difference than their prograde counterparts ($a > 0$). Consequently, counter-rotating binaries serve as marginally more sensitive probes of the magnetic environment.

\section{Conclusion}\label{sec:conc}

In this work, we have investigated the orbital dynamics and gravitational-wave signatures of extreme mass-ratio inspirals (EMRIs) in the Kerr-Bertotti-Robinson spacetime. By implementing a semi-analytic adiabatic evolution scheme driven by exact geodesic relations, we quantified the imprint of an asymptotically uniform magnetic field on the binary evolution. Our analysis complements previous mathematical studies of exact type-D solutions~\cite{Podolsky:2025tle,Wang:2025bjf} by specifically exploring their implications for gravitational-wave astronomy, and reveals three main physical effects.

First, we find a robust ``magnetic hardening'' and non-trivial topology in the gravitational-wave spectrum. Although the magnetic field pushes the ISCO to larger radii ($r_{\rm ISCO}$ increases), it simultaneously introduces geometrical confinement that requires higher orbital frequencies for circular stability, leading to a systematic blue-shift of the cutoff frequency. Furthermore, for sufficiently strong fields ($B \gtrsim 0.10$), the orbital frequency exhibits a non-monotonic evolution (``frequency turnover''), initially decreasing in the far zone before increasing in the near-horizon region. This behavior fundamentally contrasts with the monotonic chirp universally observed in vacuum black hole spacetimes.

Second, our global parameter-space scan highlights the hypersensitivity of retrograde orbits to environmental magnetic fields. We observe distinct frequency crossover phenomena; for instance, above a characteristic field strength ($B \gtrsim 0.12$ in our examples), the ISCO frequency of a counter-rotating orbit can exceed that of a nonspinning Schwarzschild black hole. More strikingly, we find that for sufficiently strong fields ($B \gtrsim 0.20$), the retrograde frequency can even surpass that of a prograde orbit. This demonstrates that sufficiently strong magnetic fields can modify---or even invert---the usual spin-frequency hierarchy at the ISCO. In a data-analysis context, neglecting such effects could introduce non-negligible biases, particularly in the inferred spin of the central black hole.

Third, these dynamical modifications manifest as substantial dephasing in the time-domain waveform. The magnetic enhancement of the orbital frequency leads to a pronounced increase in the gravitational-wave energy flux ($\dot{E}_{\rm GW} \propto \Omega_\phi^{10/3}$), driving an accelerated inspiral. Our quantitative analysis of the accumulated phase difference over the final year of inspiral demonstrates that this environmental signature is detectable by space-based detectors like LISA for magnetic fields as weak as $B \sim 10^{-4}$. Moreover, for the astrophysically realistic upper bound of $B \sim 10^{-2}$, the magnetic imprint dominates the dynamics to such an extent that the signal becomes distinguishable from vacuum waveforms within a timescale of hours, providing a highly robust observational channel.

We also briefly address the potential impact of a non-vanishing electric charge on the inspiraling body. While our analysis assumes electrically neutral test particles---a physically motivated approximation for macroscopic compact objects in EMRIs due to rapid charge neutralization by astrophysical plasmas---it is instructive to qualitatively estimate the dynamical consequences of a residual charge. Unlike the geometric curvature corrections derived in this work, which scale quadratically with the magnetic field ($\sim B^2$), the Lorentz force acting on a charged body scales linearly ($\sim qB$) and can thus dominate the dynamics if the specific charge is significant. Specifically, a repulsive Lorentz force would provide additional radial support, theoretically allowing stable circular orbits to survive at radii smaller than the neutral ISCO. Such a charge-induced inward shift would act to further enhance the frequency blue-shift signature reported here. Therefore, the geometric effects presented in this paper constitute the fundamental environmental baseline, upon which charge-dependent corrections would be superimposed.

While the present analysis provides a relativistic description of the inspiral phase within the adiabatic approximation, a comprehensive waveform template will also require modeling the merger and ringdown stages. A detailed quantitative investigation of charged particle orbits, as well as the study of quasinormal modes and perturbations in the Kerr-BR background, exceeds the scope of the present study and will be addressed in future work to enable a consistent description of the full coalescence signal.

\bmhead{Acknowledgements}

This work was supported by the National Natural Science Foundation of China under Grant No.~12305070, and the Basic Research Program of Shanxi Province under Grant Nos.~202303021222018 and 202303021221033. We thank the anonymous referee for valuable comments and suggestions that significantly deepened the discussion in this work.

\bibliography{sn-bibliography}

@article{LIGOScientific:2018mvr,
    author = "Abbott, B. P. and others",
    collaboration = "LIGO Scientific, Virgo",
    title = "{GWTC-1: A Gravitational-Wave Transient Catalog of Compact Binary Mergers Observed by LIGO and Virgo during the First and Second Observing Runs}",
    eprint = "1811.12907",
    archivePrefix = "arXiv",
    primaryClass = "astro-ph.HE",
    reportNumber = "LIGO-P1800307",
    doi = "10.1103/PhysRevX.9.031040",
    journal = "Phys. Rev. X",
    volume = "9",
    number = "3",
    pages = "031040",
    year = "2019"
}

@article{LIGOScientific:2020ibl,
    author = "Abbott, R. and others",
    collaboration = "LIGO Scientific, Virgo",
    title = "{GWTC-2: Compact Binary Coalescences Observed by LIGO and Virgo During the First Half of the Third Observing Run}",
    eprint = "2010.14527",
    archivePrefix = "arXiv",
    primaryClass = "gr-qc",
    reportNumber = "P2000061",
    doi = "10.1103/PhysRevX.11.021053",
    journal = "Phys. Rev. X",
    volume = "11",
    number = "2",
    pages = "021053",
    year = "2021"
}

@article{KAGRA:2021vkt,
    author = "Abbott, R. and others",
    collaboration = "KAGRA, VIRGO, LIGO Scientific",
    title = "{GWTC-3: Compact Binary Coalescences Observed by LIGO and Virgo during the Second Part of the Third Observing Run}",
    eprint = "2111.03606",
    archivePrefix = "arXiv",
    primaryClass = "gr-qc",
    reportNumber = "LIGO-P2000318",
    doi = "10.1103/PhysRevX.13.041039",
    journal = "Phys. Rev. X",
    volume = "13",
    number = "4",
    pages = "041039",
    year = "2023"
}

@article{EHT2019,
    author = "{Event Horizon Telescope Collaboration}",
    title = "{First M87 Event Horizon Telescope Results. I. The Shadow of the Supermassive Black Hole}",
    doi = "10.3847/2041-8213/ab0ec7",
    journal = "Astrophys. J. Lett.",
    volume = "875",
    number = "1",
    pages = "L1",
    year = "2019"
}

@article{EHT2022,
    author = "{Event Horizon Telescope Collaboration}",
    title = "{First Sagittarius A* Event Horizon Telescope Results. I. The Shadow of the Supermassive Black Hole in the Center of the Milky Way}",
    doi = "10.3847/2041-8213/ac6674",
    journal = "Astrophys. J. Lett.",
    volume = "930",
    number = "2",
    pages = "L12",
    year = "2022"
}

@article{Kerr:1963ud,
    author = "Kerr, Roy P.",
    title = "{Gravitational field of a spinning mass as an example of algebraically special metrics}",
    doi = "10.1103/PhysRevLett.11.237",
    journal = "Phys. Rev. Lett.",
    volume = "11",
    pages = "237--238",
    year = "1963"
}

@article{Melvin:1963qx,
    author = "Melvin, M. A.",
    title = "{Pure magnetic and electric geons}",
    doi = "10.1016/0031-9163(64)90801-7",
    journal = "Phys. Lett.",
    volume = "8",
    pages = "65--70",
    year = "1964"
}

@article{Ernst:1976bsr,
    author = "Ernst, Frederick J. and Wild, Walter J.",
    title = "{Kerr black holes in a magnetic universe}",
    doi = "10.1063/1.522875",
    journal = "J. Math. Phys.",
    volume = "17",
    number = "2",
    pages = "182",
    year = "1976"
}

@article{Wald:1974np,
    author = "Wald, Robert M.",
    title = "{Black hole in a uniform magnetic field}",
    doi = "10.1103/PhysRevD.10.1680",
    journal = "Phys. Rev. D",
    volume = "10",
    pages = "1680--1685",
    year = "1974"
}

@article{Podolsky:2025tle,
    author = "Podolsky, Jiri and Ovcharenko, Hryhorii",
    title = "{Kerr Black Hole in a Uniform Bertotti-Robinson Magnetic Field: An Exact Solution}",
    eprint = "2507.05199",
    archivePrefix = "arXiv",
    primaryClass = "gr-qc",
    doi = "10.1103/rfgv-ybz5",
    journal = "Phys. Rev. Lett.",
    volume = "135",
    number = "18",
    pages = "181401",
    year = "2025"
}

@article{Wang:2025bjf,
    author = "Wang, Tower",
    title = "{Innermost stable circular orbit of Kerr-Bertotti-Robinson black holes and inspirals from it: Exact solutions}",
    eprint = "2508.04684",
    archivePrefix = "arXiv",
    primaryClass = "gr-qc",
    year = "2025",
    note = "arXiv:2508.04684"
}

@article{Zeng:2025olq,
    author = "Zeng, Xiao-Xiong and Wang, Ke",
    title = "{Energy extraction from the Kerr-Bertotti-Robinson black hole via magnetic reconnection in a circular and a plunging plasma}",
    eprint = "2507.21777",
    archivePrefix = "arXiv",
    primaryClass = "gr-qc",
    doi = "10.1103/vc96-snjm",
    journal = "Phys. Rev. D",
    volume = "112",
    number = "6",
    pages = "064032",
    year = "2025"
}

@article{Wang:2025vsx,
    author = "Wang, Xinyu and Hou, Yehui and Wan, Xi and Guo, Minyong and Chen, Bin",
    title = "{Geodesics and Shadows in the Kerr-Bertotti-Robinson Black Hole Spacetime}",
    eprint = "2507.22494",
    archivePrefix = "arXiv",
    primaryClass = "gr-qc",
    year = "2025",
    note = "arXiv:2510.07914"
}

@article{Zeng:2025tji,
    author = "Zeng, Xiao-Xiong and Yang, Chen-Yu and Yu, Hao",
    title = "{Optical characteristics of the Kerr--Bertotti--Robinson black hole}",
    eprint = "2508.03020",
    archivePrefix = "arXiv",
    primaryClass = "gr-qc",
    doi = "10.1140/epjc/s10052-025-14989-y",
    journal = "Eur. Phys. J. C",
    volume = "85",
    number = "11",
    pages = "1242",
    year = "2025"
}

@article{Astorino:2025lih,
    author = "Astorino, Marco",
    title = "{Black holes in the external Bertotti-Robinson-Bonnor-Melvin electromagnetic field}",
    eprint = "2508.12908",
    archivePrefix = "arXiv",
    primaryClass = "gr-qc",
    reportNumber = "LIFT-11-4.25",
    doi = "10.1103/c5lw-53yd",
    journal = "Phys. Rev. D",
    volume = "112",
    number = "10",
    pages = "104077",
    year = "2025"
}

@article{Ali:2025beh,
    author = "Ali, Heena and Ghosh, Sushant G.",
    title = "{Parameter Estimation of Magnetised Kerr Black Holes Using Their Shadows}",
    eprint = "2508.15862",
    archivePrefix = "arXiv",
    primaryClass = "gr-qc",
    year = "2025",
    note = "arXiv:2508.15862"
}

@article{Vachher:2025jsq,
    author = "Vachher, Amnish and Kumar, Arun and Ghosh, Sushant G.",
    title = "{The influence of uniform magnetic fields on strong field gravitational lensing by Kerr black holes}",
    eprint = "2508.21100",
    archivePrefix = "arXiv",
    primaryClass = "gr-qc",
    doi = "10.1088/1475-7516/2025/11/021",
    journal = "JCAP",
    volume = "2025",
    number = "11",
    pages = "021",
    year = "2025"
}

@article{Rueda:2025lgq,
    author = "Rueda, J. A. and Ruffini, R. and Wang, Yu",
    title = "{Short GRB 090510: A magnetized neutron star binary merger leading to a black hole}",
    eprint = "2509.08172",
    archivePrefix = "arXiv",
    primaryClass = "astro-ph.HE",
    doi = "10.1016/j.jheap.2025.100464",
    journal = "JHEAp",
    volume = "50",
    pages = "100464",
    year = "2026"
}

@article{Zhang:2025ole,
    author = "Zhang, Yu-Kun and Wei, Shao-Wen",
    title = "{Effects of magnetic fields on spinning test particles orbiting Kerr-Bertotti-Robinson black holes}",
    eprint = "2510.07914",
    archivePrefix = "arXiv",
    primaryClass = "gr-qc",
    year = "2025",
    note = "arXiv:2510.07914"
}

@inproceedings{Podolsky:2025zlm,
    author = "Podolsky, Jiri",
    title = "{Various metric forms of all type D black holes and their application}",
    booktitle = "{24th International Conference on General Relativity and Gravitation (GR24) and 16th Edoardo Amaldi Conference on Gravitational Waves (Amaldi16)}",
    eprint = "2511.01029",
    archivePrefix = "arXiv",
    primaryClass = "gr-qc",
    year = "2025"
}

@inproceedings{Ovcharenko:2025qov,
    author = "Ovcharenko, Hryhorii and Podolsky, Jiri",
    title = "{A novel class of rotating black holes with non-aligned electromagnetic field}",
    booktitle = "{24th International Conference on General Relativity and Gravitation (GR24) and 16th Edoardo Amaldi Conference on Gravitational Waves (Amaldi16)}",
    eprint = "2511.04840",
    archivePrefix = "arXiv",
    primaryClass = "gr-qc",
    year = "2025"
}

@article{Ahmed:2025ril,
    author = "Ahmed, Faizuddin and Sakall{\i}, {\.I}zzet and Al-Badawi, Ahmad",
    title = "{Kerr-Bertotti-Robinson Black Holes Surrounded by a Cloud of Strings}",
    eprint = "2511.11792",
    archivePrefix = "arXiv",
    primaryClass = "gr-qc",
    year = "2025",
    note = "arXiv:2511.11792"
}

@article{Ortaggio:2025sip,
    author = "Ortaggio, Marcello",
    title = "{Einstein-Maxwell fields as solutions of Einstein gravity coupled to conformally invariant non-linear electrodynamics}",
    eprint = "2511.13665",
    archivePrefix = "arXiv",
    primaryClass = "gr-qc",
    year = "2025",
    note = "arXiv:2511.13665"
}

@article{Gray:2025lwy,
    author = "Gray, Finnian and Kubiznak, David and Ovcharenko, Hryhorii and Podolsky, Jiri",
    title = "{Hidden symmetries and separability structures of Ovcharenko-Podolsk{\'y} and conformal-to-Carter spacetimes}",
    eprint = "2511.21538",
    archivePrefix = "arXiv",
    primaryClass = "gr-qc",
    year = "2025",
    note = "arXiv:2511.21538"
}

@article{Kiselev:2002dx,
    author = "Kiselev, V. V.",
    title = "{Quintessence and black holes}",
    eprint = "gr-qc/0210040",
    archivePrefix = "arXiv",
    doi = "10.1088/0264-9381/20/6/310",
    journal = "Class. Quant. Grav.",
    volume = "20",
    pages = "1187--1198",
    year = "2003"
}

@article{Li:2019lhr,
    author = "Li, Xiang-Qian and Chen, Bo and Xing, Li-li",
    title = "{Charged Lovelock black holes in the presence of dark fluid with a nonlinear equation of state}",
    eprint = "1905.08156",
    archivePrefix = "arXiv",
    primaryClass = "gr-qc",
    doi = "10.1140/epjp/s13360-020-00231-z",
    journal = "Eur. Phys. J. Plus",
    volume = "135",
    number = "2",
    pages = "175",
    year = "2020"
}

@article{Li:2022csn,
    author = "Li, Xiang-Qian and Chen, Bo and Xing, Li-Li",
    title = "{Black holes surrounded by modified Chaplygin gas in Lovelock theory of gravity}",
    doi = "10.1016/j.aop.2022.169125",
    journal = "Annals Phys.",
    volume = "446",
    pages = "169125",
    year = "2022"
}

@article{Li:2023zfl,
    author = "Li, Xiang-Qian and Yan, Hao-Peng and Xing, Li-Li and Zhou, Shi-Wei",
    title = "{Critical behavior of AdS black holes surrounded by dark fluid with Chaplygin-like equation of state}",
    eprint = "2305.03028",
    archivePrefix = "arXiv",
    primaryClass = "gr-qc",
    doi = "10.1103/PhysRevD.107.104055",
    journal = "Phys. Rev. D",
    volume = "107",
    number = "10",
    pages = "104055",
    year = "2023"
}

@article{Letelier:1979ej,
    author = "Letelier, P. S.",
    title = "{Clouds of strings in general relativity}",
    doi = "10.1103/PhysRevD.20.1294",
    journal = "Phys. Rev. D",
    volume = "20",
    pages = "1294--1302",
    year = "1979"
}

@article{Barriola:1989hx,
    author = "Barriola, Manuel and Vilenkin, Alexander",
    title = "{Gravitational Field of a Global Monopole}",
    reportNumber = "TUTP-89-4",
    doi = "10.1103/PhysRevLett.63.341",
    journal = "Phys. Rev. Lett.",
    volume = "63",
    pages = "341",
    year = "1989"
}

@article{Detweiler:1980uk,
    author = "Detweiler, Steven L.",
    title = "{Klein-Gordon equation and rotating black holes}",
    doi = "10.1103/PhysRevD.22.2323",
    journal = "Phys. Rev. D",
    volume = "22",
    pages = "2323--2326",
    year = "1980"
}

@article{Zouros:1979iw,
    author = "Zouros, T. J. M. and Eardley, D. M.",
    title = "{Instabilities of massive scalar perturbations of a rotating black hole}",
    doi = "10.1016/0003-4916(79)90237-9",
    journal = "Annals Phys.",
    volume = "118",
    pages = "139--155",
    year = "1979"
}

@book{Chandrasekhar:1985kt,
    author = "Chandrasekhar, Subrahmanyan",
    title = "{The mathematical theory of black holes}",
    isbn = "978-0-19-850370-5",
    year = "1985"
}

@article{Zi:2023qfk,
    author = "Zi, Tieguang and Li, Peng-Cheng",
    title = "{Gravitational waves from extreme-mass-ratio inspirals in the semiclassical gravity spacetime}",
    eprint = "2311.07279",
    archivePrefix = "arXiv",
    primaryClass = "gr-qc",
    doi = "10.1103/PhysRevD.109.064089",
    journal = "Phys. Rev. D",
    volume = "109",
    number = "6",
    pages = "064089",
    year = "2024"
}

\end{document}